\documentclass[a4paper, 11pt]{article}
\usepackage{amsmath, amssymb, amsthm}
\usepackage{graphicx}
\usepackage{latexsym}

\def\b{\begin{eqnarray}}
\def\e{\end{eqnarray}}
\def\n{\noindent}

\newtheorem{theorem}{Theorem}

\newtheorem{definition}{Definition}
\newtheorem*{acknowledgments}{Acknowledgments}
\begin{document}

\begin{center}

{\LARGE\textbf{On the Integrability of a Class of Nonlinear
Dispersive Wave Equations
\\}} \vspace {10mm} \vspace{1mm} \noindent

{\large \bf Rossen I. Ivanov}\footnote{On leave from the Institute
for Nuclear Research and Nuclear Energy, Bulgarian Academy of
Sciences, Sofia, Bulgaria.} \vskip1cm \n \hskip-.3cm
\begin{tabular}{c}
\hskip-1cm {\it School of Mathematics, Trinity College Dublin,}
\\ {\it Dublin 2, Ireland} \\
{\it Tel:  + 353 - 1 - 608 2898 }\\{\it  Fax:  + 353 - 1- 608 2282} \\
\\ {\it e-mail: ivanovr@tcd.ie}
\\
\hskip-.8cm
\end{tabular}
\vskip1cm
\end{center}


\begin{abstract}
\noindent We investigate the integrability of a class of 1+1
dimensional models describing nonlinear dispersive waves in
continuous media, e.g. cylindrical compressible hyperelastic rods,
shallow water waves, etc. The only completely integrable cases
coincide with the Camassa-Holm and Degasperis-Procesi equations.
\end{abstract}

\section{Introduction}

In this letter we investigate the integrability of the nonlinear
equation
\begin{equation}\label{eq1}
 u_{t}-u_{xxt}+\partial_{x}g[u]=\nu u_{x}u_{xx}+\gamma
 uu_{xxx}, 
\end{equation}
where
\begin{equation}\label{eq2}
  g[u]=\kappa u+\alpha u^{2} +\beta u ^{3}
\end{equation}
and $\alpha$, $\beta$, $\gamma$, $\kappa$, $\nu$ are constant
parameters. The symmetries of (\ref{eq1}) for specific choices of
the parameters are studied in \cite{CMP97}.

The case $\kappa=0$, $\alpha=3/2$, $\beta=0$, $\nu=2\gamma$ and
$\gamma$ an arbitrary real parameter has been recently studied as a
model, describing nonlinear dispersive waves in cylindrical
compressible hyperelastic rods \cite{Dai00, Dai98} -- see also
\cite{CS00, Y03}. The physical parameters of various compressible
materials put $\gamma$ in the range from -29.4760 to 3.4174
\cite{Dai00, Dai98}.

Other important cases of (\ref{eq1}) are:

Camassa-Holm (CH) equation \cite{CH93, FF81}
\begin{equation}\label{eq18}
 u_{t}-u_{xxt}+\kappa u_{x} +3uu_{x}=2u_{x}u_{xx}+uu_{xxx},
\end{equation}
$\kappa - \text{arbitrary (real)}$, describing the unidirectional
propagation of shallow water waves over a flat bottom \cite{CH93,
J02}. CH is a completely integrable equation \cite{BBS98, CM99, C01,
L02}, describing permanent and breaking waves \cite{CE98, C00}. The
solitary waves of CH are smooth if $\kappa
> 0$ and peaked if $\kappa = 0$ \cite{CH93, LOR99}. Integrable generalizations of CH with
higher order terms are derived in \cite{F96}.

Degasperis-Procesi (DP) equation \cite{DP99}:
\begin{equation}\label{eq18a}
 u_{t}-u_{xxt}+\kappa u_{x}+4uu_{x}=3u_{x}u_{xx}+uu_{xxx},
\end{equation}
$\kappa - \text{arbitrary (real)}$, is another completely integrable
equation of this class. It is also known to possess (multi)peakon
solutions if $\kappa = 0$ \cite{DHH02, HJPW03}.

CH and DP equations are particular cases of the $b$-family
\begin{equation}\label{eq18b}
 u_{t}-u_{xxt}+(b+1)uu_{x}=bu_{x}u_{xx}+uu_{xxx},
\end{equation}
which possesses multipeakon  solutions for any real $b$
\cite{DHH02}.

 Fornberg-Whitham (FW) equation \cite{W67}
\begin{equation}\label{eq18d}
 u_{t}-u_{xxt}+ u_{x}+uu_{x}=3u_{x}u_{xx}+uu_{xxx}
\end{equation}
appeared in the study of the qualitative behaviors of wave-breaking.

The regularized long-wave (RLW) or BBM equation \cite{BBM72}
\begin{equation}\label{eq18e}
 u_{t}-u_{xxt}+ u_{x}+uu_{x}=0
\end{equation}
and the modified BBM equation
\begin{equation}\label{eq18f}
u_{t}-u_{xxt}+ u_{x}+(u^{3})_{x}=0
\end{equation}
are not completely integrable, although they have three nontrivial
independent integrals~\cite{O79}.

In what follows we will demonstrate that the only completely
integrable representatives of the class (\ref{eq1}) are CH and DP
equations (\ref{eq18}), (\ref{eq18a}).


In our analysis we will use the integrability check developed in
 \cite{MN02, SJ98, OJ00}. This perturbative method can be briefly outlined as
 follows. Consider the evolution partial differential equation
\begin{equation}\label{eq3}
 u_{t}=F_{1}[u]+F_{2}[u]+F_{3}[u]+\ldots
\end{equation}
where $F_{k}[u]$ is a homogeneous differential polynomial, i.e. a
polynomial of variables $u$, $u_{x}$, $u_{xx}$, ..., $\partial
^{n}_{x}u$ with complex constant coefficients, satisfying the
condition
\begin{equation}
 F_{k}[\lambda u]=\lambda ^{k}F_{k}[u], \qquad \lambda \in  \mathbb{C}. \nonumber
\end{equation}
The linear part is $F_{1}[u]=L(u)$, where $L$ is a linear
differential operator of order two or higher. The representation
(\ref{eq3}) can be put into correspondence to a symbolic expression
of the form

\begin{equation}\label{eq5}
 u_{t}=u\omega (\xi_{1})+\frac{u^{2}}{2}a_{1}(\xi _{1}, \xi _{2})+ \frac{u^{3}}{3}a_{2}(\xi _{1}, \xi _{2}, \xi
 _{3})+\ldots=F
\end{equation}
where $\omega (\xi_{1})$ is a polynomial of degree 2 or higher and
$a_{k}(\xi _{1}, \xi _{2},\ldots \xi _{k+1})$ are symmetric
polynomials. Each of these polynomials is related to the Fourier
image of the corresponding $F_{k}[u]$ and can be obtained through a
simple procedure, described e.g. in \cite{MN02}. Each differential
monomial $u^{n_{0}}u_{x}^{n_{1}}\ldots(\partial_{x}^{q}u)^{n_{q}}$
is represented by a symbol
\begin{equation}
u^{m}\langle \xi_{1}^{0}\ldots
\xi_{n_{0}}^{0}\xi_{n_{0}+1}^{1}\ldots
\xi_{n_{0}+n_{1}}^{1}\xi_{n_{0}+n_{1}+1}^{2}\ldots
\xi_{n_{0}+n_{1}+n_{2}}^{2}\ldots \xi_{m}^{q}\rangle \nonumber
\end{equation}
where $m=n_{0}+n_{1}+\ldots n_{q}$ and the brackets $\langle\rangle$
denote symmetrization over all arguments $\xi_{k}$ (i.e.
symmetrization with respect to the group of permutations of $m$
elements $S_{m}$):
\begin{equation}
\langle f(\xi_{1},\xi_{2},\ldots,
\xi_{n})\rangle=\frac{1}{m!}\sum_{\sigma\in S_{m}}
f(\xi_{\sigma(1)},\xi_{\sigma(2)},\ldots, \xi_{\sigma(n)}) \nonumber
\end{equation}
Also, for any function $F$ (\ref{eq5}) there exists a formal
recursion operator
\begin{equation}\label{eq6}
 \Lambda = \eta +u \phi_{1}(\xi_{1},\eta)+u^{2}\phi_{2}(\xi _{1}, \xi _{2},\eta)+ \ldots
\end{equation}
where the coefficients $\phi_{m}(\xi _{1}, \xi _{2},\ldots \xi _{m},
\eta)$ can be determined recursively:

\begin{subequations} \label{eq7}
\begin{gather}
  \phi_{1}(\xi_{1},\eta) = N^{\omega}(\xi_{1},\eta)\xi_{1}a_{1}(\xi_{1},\eta)      \label{eq6a}\\
  \phi_{m}(\xi _{1}, \xi _{2},\ldots \xi _{m},\eta) = N^{\omega}(\xi _{1}, \xi _{2},\ldots \xi _{m},\eta)\Big \{
  (\sum_{p=1}^m \xi _{p})a_{m}(\xi _{1}, \xi _{2},\ldots \xi
  _{m},\eta) + \nonumber \\
  +\sum_{n=1}^{m-1}\Big \langle \frac{n}{m-n+1}\phi_{n}(\xi _{1},\ldots \xi _{n-1},\xi _{n}+\ldots+\xi _{m},\eta)
   a_{m-n}(\xi _{n},\ldots \xi _{m})  + \nonumber \\
+\phi_{n}(\xi _{1},\ldots \xi _{n},\eta+\xi _{n+1}+\ldots+\xi_{m})
   a_{m-n}(\xi _{n+1},\ldots \xi _{m},\eta)  - \nonumber \\
   -\phi_{n}(\xi _{1},\ldots \xi _{n},\eta)
   a_{m-n}(\xi _{n+1},\ldots \xi _{m},\eta+\xi _{1}+\ldots+\xi_{n})\Big \rangle\Big\}    \label{eq6c}
\end{gather}
\end{subequations}
with
\begin{equation}\label{eq8}
 N^{\omega}(\xi _{1}, \xi _{2},\ldots \xi _{m}) = \Big(\omega(\sum_{n=1}^{m}\xi _{n})-\sum_{n=1}^{m}\omega (\xi_{n})\Big)^{-1}
\end{equation}
and the symbols $\langle \rangle$ denote symmetrization with respect
to $\xi _{1}, \xi _{2},\ldots \xi _{m}$, (the symbol $\eta$ is not
included in the symmetrization). Before formulating the
integrability criterion it is necessary to introduce the following
\begin{definition}
The function $b_{m}(\xi _{1}, \xi _{2},\ldots \xi _{m},\eta)$ ,
$m\geq 1$ is called local if all coefficients
$b_{mn}(\xi_{1},\xi_{2},\ldots \xi _{m})$, $n=n_{s},n_{s+1},\ldots$
of its expansion as $\eta\rightarrow \infty$
\begin{equation}\label{eq9}
 b_{m}(\xi _{1}, \xi _{2},\ldots \xi _{m},\eta) =\sum_{n=n_{s}}^{\infty}b_{mn}(\xi _{1}, \xi _{2},\ldots \xi _{m})\eta^{-n}
\end{equation}
are symmetric polynomials.
\end{definition}

Now the integrability criterion can be summarized as follows
\cite{MN02}:
\begin{theorem} \label{th1}
  The complete integrability of the equation (\ref{eq3}), i.e. the existence of
an infinite hierarchy of local symmetries or conservation laws,
implies that all the coefficients (\ref{eq7}) of the formal
recursion operator(\ref{eq6}) are local.
\end{theorem}

\section{The integrability test}

After shifting $u\rightarrow -(u+\delta)$ and $x \rightarrow
x-\lambda t$ where $\delta$ and $\lambda$ are arbitrary constants,
the equation (\ref{eq1}) can be written in the form
\begin{equation}\label{eq10}
 u_{t}=(1-\partial ^{2}_{x} )^{-1}\Big( K u_{x}+
 Bu_{xxx}+Cuu_{x}+Au^{2}u_{x}-\nu u_{x}u_{xx}-\gamma uu_{xxx}\Big)
\end{equation}
where the new constants $A$, $B$, $C$ and $K$ are related to the old
ones as follows:

\begin{subequations} \label{eq11}
\begin{gather}
  A=-3\beta      \label{eq11a}\\
  B=\lambda -\gamma \delta   \label{eq11b} \\
C=2\alpha-6\beta \delta   \label{eq11c} \\
K=2\alpha \delta -\kappa-3\beta\delta^{2} -\lambda \label{eq11d}
\end{gather}
\end{subequations}
Since the linear part of the equation must contain second derivative
or higher, the applicability of the test requires $B\neq0$, i.e.
$\lambda\neq\gamma\delta$ which always can be achieved by a proper
choice of the arbitrary constant $\lambda$.

 The symbolic representation of the operator $(1-\partial
^{2}_{x} )^{-1}$ is $\frac{1}{1-\eta ^{2}}$ and the symbol,
corresponding to $(1-\partial ^{2}_{x} )^{-1}F_{k}[u]$ is
$\frac{u^{k}}{k}\frac{a_{k-1}(\xi _{1}, \xi _{2},\ldots \xi
_{k})}{1-(\xi _{1}+ \xi _{2}+\ldots +\xi _{k})^{2}}$, where
$\frac{u^{k}}{k}a_{k-1}(\xi _{1}, \xi _{2},\ldots \xi _{k})$ is the
symbol corresponding to $F_{k}[u]$; see \cite{MN02} for details.
Moreover, Theorem \ref{th1} can be applied in this case as well.
Therefore, the equation (\ref{eq10}) can be represented in the form
(\ref{eq5}) with
\begin{subequations} \label{eq12}
\begin{gather}
\omega(\xi_{1})=\frac{K\xi_{1}+B\xi_{1}^{3}}{1-\xi_{1}^{2}}      \label{eq12a}\\
a_{1}(\xi _{1}, \xi _{2})=\frac{C(\xi _{1}+\xi _{2})-\nu \xi _{1}\xi
_{2}(\xi _{1}+\xi _{2})-
\gamma (\xi _{1}^{3}+\xi _{2}^{3}) }{1-(\xi _{1}+\xi _{2})^{2}} \label{eq12b} \\
 a_{2}(\xi _{1}, \xi _{2},\xi _{3})=\frac{A(\xi _{1}+\xi _{2}+\xi _{3})}{1-(\xi _{1}+\xi _{2}+\xi _{3})^{2}} \label{eq12c}
  \end{gather}
\end{subequations}
Then from (\ref{eq7}):
\begin{subequations} \label{eq13}
\begin{gather}
\phi_{1}(\xi_{1},\eta)=\frac{(1-\xi_{1}^{2})(1-\eta^{2})\Big(-C+\gamma
\xi_{1}^{2}+(\nu-\gamma) \xi_{1}\eta +\gamma \eta^{2}\Big)}
 {(B+K)\eta(-3+ \xi_{1}^{2}+\xi_{1}\eta +\eta^{2})}      \label{eq13a}\\
   \phi_{2}(\xi_{1},\xi_{2},\eta)=\Phi_{21}(\xi_{1},\xi_{2})\eta +
\Phi_{20}(\xi_{1},\xi_{2})+\Phi_{2,-1}(\xi_{1},\xi_{2})\eta
^{-1}\nonumber \\ +\Phi_{2,-2}(\xi_{1},\xi_{2})\eta ^{-2} +\ldots
\label{eq13b}
  \end{gather}
\end{subequations}

All coefficients in the expansion of $\phi_{1}(\xi_{1},\eta)$
(\ref{eq13a}) with respect to $\eta$ are polynomials on $\xi_{1}$
and therefore there are no obstacles to the integrability of
(\ref{eq10}). However, the expansion of
$\phi_{2}(\xi_{1},\xi_{2},\eta)$ (\ref{eq13b}) may contain, in
general, non-polynomial contributions.

Let us start with the case $\gamma \neq0$. Then
\begin{equation}\label{eq14}
 \Phi_{21}(\xi_{1},\xi_{2})=\gamma\frac{(1-\xi_{1}^{2})(1-\xi_{2}^{2})\Big(-C+(\gamma+\nu) \xi_{1} \xi_{2}\Big)}
 {(B+K)^{2}(1- \xi_{1}\xi_{2})}
\end{equation}
is a polynomial iff
\begin{equation}\label{eq15}
 C=\gamma +\nu.
\end{equation}
Then
\begin{equation}\label{eq16}
 \Phi_{2,-1}(\xi_{1},\xi_{2})=\frac{(1-\xi_{1}^{2})(1-\xi_{2}^{2})P_{1}(\xi_{1}, \xi_{2})}
 {(B+K)^{2}(1- \xi_{1}\xi_{2})}
\end{equation}
where
\begin{subequations}
\begin{gather} \nonumber
P_{1}(\xi_{1}, \xi_{2})=-A(B+K)+ (\xi_{1}\xi_{2}-1)\times \\
\nonumber \times
\Big(\gamma^{2}(2\xi_{1}+\xi_{2})(\xi_{1}+2\xi_{2})+\gamma \nu
(1+\xi_{1}^{2}+\xi_{1}\xi_{2}+\xi_{2}^{2})-\nu^{2}(1+(\xi_{1}+\xi_{2})^{2})\Big).
\end{gather}
 \end{subequations}
$\Phi_{2,-1}(\xi_{1},\xi_{2})$ is a polynomial iff $A=0$. From
(\ref{eq11}), (\ref{eq15}) we get
\begin{equation}\label{eq17}
 \beta=0, \qquad \alpha=\frac{\gamma+\nu}{2}.
\end{equation}
With $\alpha$ and $\beta$ as in (\ref{eq17}), the term
$\Phi_{2,-3}(\xi_{1},\xi_{2})$ has the form
\begin{equation}\label{eq19}
 \Phi_{2,-3}(\xi_{1},\xi_{2})=\frac{P_{2}(\xi_{1}, \xi_{2})}
 {(B+K)^{2}(1- \xi_{1}\xi_{2})}
\end{equation}
where $P_{2}(\xi_{1}, \xi_{2})$ is a symmetric polynomial. The
polynomial remainder of the division of $P_{2}(\xi_{1}, \xi_{2})$
with $1-\xi_{1}\xi_{2}$ (e.g. if $\xi_{2}$ is treated as a constant,
and $\xi_{1}$ as a polynomial variable) is proportional to the
factor $6\gamma^{2}-5\gamma \nu +\nu^{2}$. Thus, for complete
integrability it is necessary
\begin{equation}\label{eq20}
 6\gamma^{2}-5\gamma \nu +\nu^{2}=0.
\end{equation}
There are two nonzero solutions of (\ref{eq20}): $\nu =2\gamma$ and
$\nu =3\gamma$. From (\ref{eq17}), $\alpha=\frac{3\gamma}{2}$ and
$\alpha=2\gamma$  in these two cases correspondingly. The
requirement $B+K\neq 0 $ (\ref{eq13a}) or $\kappa\neq \nu \delta$
can be achieved for suitable $\delta$, if $\nu\neq0$, or even for
$\nu=0$ if $\kappa\neq0$. The test is inconclusive if
$\kappa=\nu=0$, which corresponds to the equation (\ref{eq18b}) with
$b=0$. This case is not integrable, although it admits a Hamiltonian
formulation \cite{DHH02}.

Without loss of generality one can choose now $\gamma=1$ (e.g. after
rescaling of $t$), which gives precisely the integrable Camassa-Holm
and Degasperis-Procesi equations (\ref{eq18}), (\ref{eq18a}).
%

Now suppose $\gamma=0$, $\nu\neq0$. In this case
\begin{equation}\label{eq21}
 \Phi_{20}(\xi_{1},\xi_{2})=\nu\frac{(1-\xi_{1}^{2})(1-\xi_{2}^{2})(\xi_{1}+\xi_{2})(C-\nu \xi_{1} \xi_{2})}
 {(B+K)^{2}(1- \xi_{1}\xi_{2})}
\end{equation}
is a polynomial iff $C=\nu$. Then
\begin{equation}\label{eq22}
 \Phi_{2,-1}(\xi_{1},\xi_{2})=\nu\frac{(1\!\!-\!\!\xi_{1}^{2})(1\!\!-\!\!\xi_{2}^{2})\Big(\!\!-\!\!A(B+K)
 +\nu^{2}(1\!\!-\!\!\xi_{1}\xi_{2})(1+( \xi_{1}\!\!+ \!\!\xi_{2})^{2})\Big)}
 {(B+K)^{2}(1- \xi_{1}\xi_{2})}
\end{equation}
is a polynomial iff $A=0$, i.e. $\beta=0$. If $C=\nu$ and $\beta=0$
(i.e. $\alpha=\nu/2)$ a further computation gives
\begin{equation}\label{eq23}
 \Phi_{2,-3}(\xi_{1},\xi_{2})=-\nu^{2}\frac{(1-\xi_{1}^{2})(1-\xi_{2}^{2})P_{3}(\xi_{1}, \xi_{2})}
 {(B+K)^{2}(1- \xi_{1}\xi_{2})}
\end{equation}
where
\begin{subequations}
\begin{gather} \nonumber
 P_{3}(\xi_{1},\xi_{2})=3\!\!+\!\!7\xi_{1}^{2}+7\xi_{2}^{2}\!\!-\!\!\xi_{1}^{4}\!\!-\!\!\xi_{2}^{4}+12
 \xi_{1}\xi_{2}-29\xi_{1}^{2}\xi_{2}^{2}+8\xi_{1}^{3}\xi_{2}^{3}+\xi_{1}^{5}\xi_{2}+\xi_{1}\xi_{2}^{5} +\\ \nonumber
 + 8\xi_{1}^{2}\xi_{2}^{4}+8\xi_{1}^{4}\xi_{2}^{2}-12\xi_{1}^{3}\xi_{2}-12\xi_{1}\xi_{2}^{3}.
 \end{gather}
 \end{subequations}
Therefore $\Phi_{2,-3}$ (\ref{eq23}) is not a polynomial for
$\nu\neq0$.  Note that the restriction $B+K\neq 0 $ (\ref{eq13a}) is
again secured by the choice $\delta\neq\kappa/\nu$. Thus, if
$\gamma=0$ and $\nu\neq0$ no completely integrable equations emerge.

Finally, let us take $\gamma=\nu=0$. In this case
\begin{equation}\label{eq24}
 \Phi_{2,-1}(\xi_{1},\xi_{2})=\frac{\Big( A(B+K)-C^{2}\Big)(1-\xi_{1}^{2})(1-\xi_{2}^{2})}
 {(B+K)^{2}(1- \xi_{1}\xi_{2})}
\end{equation}
is a polynomial iff $C^{2}=A(B+K)$. But then
\begin{equation}\label{eq26}
 \Phi_{2,-3}(\xi_{1},\xi_{2})=\frac{A(1-\xi_{1}^{2})(1-\xi_{2}^{2})(1-4\xi_{1}\xi_{2})}
 {(B+K)(1- \xi_{1}\xi_{2})}
\end{equation}
is apparently not a polynomial if $A\neq0$, i.e. $\beta\neq0$  ( if
$\beta\neq0$, $B+K=2\alpha\delta-\kappa-3\beta\delta^{2}$ can be
arranged to be nonzero by a proper choice of $\delta$). Therefore,
the only possibility, leading to an integrable equation could be
$A=0$. Then it is obvious that for $C\neq0$ (\ref{eq24}) is not a
polynomial (in this case $B+K=C\delta-\kappa$,
$\delta\neq\kappa/C$). Thus for integrability it is necessary
$A=C=0$ but then the equation (\ref{eq10}) becomes linear.

Therefore, the only nonlinear completely integrable representatives
of the class (\ref{eq1}) are the Camassa-Holm and Degasperis-Procesi
equations (\ref{eq18}), (\ref{eq18a}).

\begin{acknowledgments}
The author is grateful to Prof. A. Constantin for helpful
discussions and to an anonymous referee for important comments and
suggestions. The author also acknowledges funding from the Science
Foundation Ireland, Grant 04/BR6/M0042.
\end{acknowledgments}

\label{lastpage}

\begin{thebibliography}{99}
\small

\bibitem{BBS98}
Beals R, Sattinger D and Szmigielski J, Acoustic Scattering and the
Extended Korteweg-de Vries Hierarchy, {\it Adv. Math.} {\bf 140}
(1998), 190--206.

\bibitem{BBM72}
Benjamin T, Bona J L  and Mahony J J, Model Equations for Long Waves
in Nonlinear Dispersive Systems, {\it Philos. Trans. R. Soc. A} {\bf
272} (1972), 47--78.

\bibitem{CH93}
Camassa R and Holm D, An Integrable Shallow Water Equation with
Peaked Solitons, {\it Phys. Rev. Lett.} {\bf 71} (1993), 1661--1664.

\bibitem{CMP97}
Clarkson P A, Mansfield E L and Priestley T J, Symmetries of a Class
of Nonlinear Third Order Partial Differential Equations, {\it Math.
Comput. Modelling} {\bf 25} (1997), 195--212.

\bibitem{C00}
Constantin A, Existence of permanent and breathing waves for a
shallow water equation: a geometric approach, {\it Ann. Inst.
Fourier} {\bf 50} (2000), 321--362.

\bibitem{C01}
Constantin A, On the Scattering Problem for the Camassa-Holm
Equation, {\it Proc. R. Soc. A} {\bf 457} (2001), 953--970.

\bibitem{CE98}
Constantin A and Escher J, Wave Breaking for Nonlinear Nonlocal
Shallow Water Equations, {\it Acta Math.} {\bf 181} (1998),
229--243.

\bibitem{CM99}
Constantin A and McKean H P, A Shallow Water Equation on the Circle,
{\it Commun. Pure Appl. Math.} {\bf 52} (1999), 949--982.

\bibitem{CS00}
Constantin A and Strauss W A, Stability of a Class of Solitary Waves
in Compressible Elastic Rods, {\it Phys. Lett. A} {\bf 270} (2000),
140--148.

\bibitem{Dai98}
Dai H-H, Model Equations for Nonlinear Dispersive Waves in a
Compressible Mooney-Rivlin Rod, {\it Acta Mech.} {\bf 127} (1998),
193--207.

\bibitem{Dai00}
Dai H-H and Huo Y, Solitary Shock Waves and Other Travelling Waves
in a General Compressible Hyperelastic Rod, {\it Proc. R. Soc. A}
{\bf 456} (2000), 331--363.

\bibitem{DP99}
Degasperis A and Procesi M, Asymptotic Integrability, in Symmetry
and Perturbation Theory, Editors: Degasperis~A and Gaeta~G, World
Scientific, Singapore, 1999, 23--37.

\bibitem{DHH02}
Degasperis A, Holm D D and Hone A N W, A New Integrable Equation
with Peakon Solutions,{\it Theor. Math. Phys.} {\bf 133} (2002),
1463--1474.

\bibitem{FF81}
Fokas A and Fuchssteiner B, Symplectic Structures, their B\"acklund
Transformation and Hereditary Symmetries, {\it Physica D} {\bf 4}
(1981), 821--831.

\bibitem{F96}
Fuchssteiner B, Some Tricks from the Symmetry-Toolbox for Nonlinear
Equations: Generalisations of the Camassa-Holm Equation, {\it
Physica D} {\bf 105} (1996), 229--242.

\bibitem{HJPW03}
Hone A N W and Jing Ping Wang, Prolongation Algebras and Hamiltonian
Operators for Peakon Equations, {\it Inverse Probl.} {\bf 19}
(2003), 129--145.

\bibitem{J02}
Johnson R~S, Camassa-Holm, Korteweg-de Vries and Related Models for
Water Waves, {\it J. Fluid. Mech.} {\bf 457} (2002), 63--82.

\bibitem{L02} Lenells J, The Scattering Approach for the Camassa-Holm Equation, {\it J. Nonlinear
Math. Phys.} {\bf 9} (2002), 389--393.

\bibitem{LOR99} Li Y~A, Olver P~J and Rosenau P, Non-analytic Solutions of Nonlinear Wave Models,
in Nonlinear Theory of Generalised Functions: Vienna 1997, {\it
Chapman $\&$ Hall/CRC Res. Notes Math.} {\bf 401}, Chapman $\&$
Hall/CRC, Boca Raton, FL, 1999, 129--145.

\bibitem{MN02}
Mikhailov A~V and Novikov V~S, Perturbative Symmetry Approach, {\it
J. Phys. A} {\bf 35} (2002), 4775--4790.

\bibitem{OJ00} Olver P and Jing Ping Wang, Classification of Integrable
One-component Systems on Associative Algebras, {\it Proc. London
Math. Soc.} {\bf 81} (2000), 566--586.

\bibitem{O79} Olver P, Euler Operators and Conservation Laws of the BBM Equation, {\it Math. Proc. Camb.
Phil. Soc.} {\bf 85} (1979), 143--160.

\bibitem{SJ98}
Sanders J and Jing Ping Wang, On the Integrability of Homogenious
Scalar Evolution Equations, {\it J. Diff. Eq.} {\bf 147} (1998),
410--434.

\bibitem{W67}
Whitham G, Variational Methods and Applications to Water Waves, {\it
Proc. R. Soc. A} {\bf 299} (1967), 6--25.

\bibitem{Y03} Yin Z, On the Cauchy Problem for a Nonlinearly Dispersive Wave Equation, {\it J. Nonlinear
Math. Phys.} {\bf 10} (2003), 10--15.








\end{thebibliography}
\end{document}